\documentclass{aa}
\usepackage{epsf}
\usepackage{graphicx}
\begin{document}
   \title{On the progenitor of V838 Monocerotis}

\author{R. Tylenda\inst{1,3}
      \and N. Soker\inst{2}
      \and R. Szczerba\inst{1}}

\offprints{R. Tylenda, \email{tylenda@ncac.torun.pl}}
\institute{Department for Astrophysics, N.Copernicus Astronomical Center,
Rabia\'nska 8, 87-100 Toru\'n, Poland\\ \email{tylenda@ncac.torun.pl,
szczerba@ncac.torun.pl}
\and
Department of Physics, Technion-Israel Institute of
Technology,  32000 Haifa, Israel\\ \email{soker@physics.technion.ac.il}
\and
Centre for Astronomy, N. Copernicus University, 87-100 Toru\'n, Poland}

\date{Received   }

\abstract{
We summarize and analyze the available observational
data on the progenitor and the enviroment of V838~Mon.
From the available photometric data for the progenitor of
V838~Mon we exclude the possibility that the object before eruption
was an evolved red giant star (AGB or RGB star).
We find that
most likely it was a main sequence or pre-main sequence star of $\sim 5-10 M_\odot$.
From the light echo structure and evolution we conclude that the reflecting dust
is of interstellar nature rather than blown by V838~Mon in the past.
We discuss the IRAS and CO data for interstellar medium observed near
the position of V838~Mon. Several interstellar molecular regions have 
radial velocities similar to that of V838~Mon, so dust seen in the light echo
might be related to one of them.

   }

\maketitle

\section{Introduction \label{int}}

V838 Mon is a star caught in eruption at the beginning of
January 2002 (Brown \cite{brown}).
The eruption, as observed in optical wavelengths, 
lasted about three months, and was composed
of two or three major peaks.
After developing an A-F supergiant spectrum
at the optical maximum at the beginning of February,
the object showed a general tendency to evolve to lower effective
temperatures. In April~2002 it almost disappeared from the optical but remained
very bright in infrared becoming one of the coolest M-type supergiants yet
observed. Detailed descriptions of the spectral and
photometric evolution of V838~Mon can be found in a number of papers
including Munari et~al. (\cite{munari}), Kimeswenger et~al. (\cite{kimes}),
Kolev et~al. (\cite{kolev}),
Osiwa\l{}a et~al. (\cite{osiwal}), Wisniewski et~al. (\cite{wisnia}), Crause
et~al. (\cite{lisa03}) and Kipper et~al. (\cite{kipper}).

The nature of the V838~Mon eruption is enigmatic.
As discussed by Soker \& Tylenda (\cite{soktyl}), thermonuclear models
(classical nova, He-shell flash) seem to
be unable to explain this type of eruption.
Therefore other mechanisms such as a stellar merger model
(Soker \& Tylenda \cite{soktyl}) or a giant swallowing planets scenario
(Retter \& Marom \cite{retmar}) have been proposed.

The global fading of V838~Mon in optical after outburst has enabled us
to discover a faint hot continuum in short wavelengths (Desidera \& Munari
\cite{desmun}; Wagner \& Starrfield \cite{wagner}) later classified as
coming from a normal B3~V star (Munari et al. \cite{mundes}). 
This strongly suggests that
V838~Mon is a binary system which can be an important fact for identifying
the outburst mechanism.

V838 Mon has received significant publicity due to its light echo,
which was discovered shortly after the main eruption in
February 2002 (Henden et al. \cite{henden}), and was
seen in images by the HST (Bond et al. \cite{bond}).
The light echo was used, e.g. in Bond et al. (\cite{bond}, \cite{bond04}),
to claim that the
echoing matter was ejected by V838 Mon in previous eruptions.
This conclusion was disputed by Tylenda (\cite{tyl}), who
examined the evolution of the light echo and
concluded that the dust illuminated by the light echo
was of interstellar origin rather than produced by mass loss from
V838 Mon in the past.

van~Loon et~al. (2004) argue
that there are multiple shells around V838 Mon, which were
ejected by V838 Mon in previous eruptions. Hence they reason that
prior to eruption V838 Mon was an asymptotic giant branch 
(AGB) star.
These authors have also analyzed the light echo
with more recent observations than in Tylenda (\cite{tyl}), and argue that
the echoing dust was ejected by V838 Mon in past eruptions.
 
In the present paper we collect and discuss the data available
on the progenitor of V838~Mon. This includes the archival photometric measurements
done in the optical and infrared before 2002, results of analysis of the evolution
of the light echo after the eruption, as well as
available data on regions of interstellar matter (ISM) near the position of
V838~Mon. In the case of erupting stars, conclusions drawn from the progenitor
usually are very important for constraining the mechanism of the eruption.
An analysis of the observational data for V838~Mon during and after its eruption
is done in another paper (Tylenda \cite{tyl05}).

\section{The analysis of the photometric data \label{phot}}

Table~\ref{mag} lists the photometric results for V838~Mon prior to its outburst.
Columns (1) and (2) give the names and the effective wavelengths of the
photometric bands. The magnitudes and the error estimates in column (3) are
from the references given in the last column.
Optical magnitudes have been taken from Kimeswenger et~al.
(\cite{kimes}) and Goranskij et~al. (\cite{goran}). 
Munari et~al. (\cite{munari}, \cite{munhen})
have also estimated magnitudes of the V838~Mon progenitor. However, the results
given in these two references differ by $\sim 1$~mag. We do not take them into account
as it is not clear
what caused such large differences (Munari et~al. \cite{munhen} do not comment
on this). However, if one takes mean values from
Munari et~al. (\cite{munari}, \cite{munhen}) they do not significantly differ from
those quoted in Table~{\ref{mag}.
The object has also been observed in infrared surveys. $JHK$ magnitudes
can be found in the 2MASS data while the DENIS experiment measured
the $IJK$ bands. 

\begin{table}
\centering
\caption{Photometry of the V838~Mon progenitor}
\label{mag}
\begin{tabular}{l c c l}
\hline
  Band       &  $\lambda_{eff}$($\mu$m)  & Magnitude & Reference\\
\hline
  $B$        &  0.44  &  15.87$\pm$0.10  & Kimeswenger et al. \cite{kimes} \\
  $B$        &  0.44  &  15.81$\pm$0.06  & Goranskij et~al. \cite{goran}  \\
  $R_c$      &  0.65  &  14.84$\pm$0.06  & Goranskij et~al. \cite{goran}  \\
  $R$        &  0.71  &  14.56$\pm$0.10  & Kimeswenger et al. \cite{kimes} \\
  $I_c$      &  0.80  &  14.27$\pm$0.03  & Goranskij et~al. \cite{goran}  \\
  $I_{Gunn}$ &  0.82  &  14.51$\pm$0.03  & DENIS \\
  $J$        &  1.25  &  13.86$\pm$0.11  & DENIS \\
  $J$        &  1.25  &  13.87$\pm$0.05  & 2MASS \\
  $H$        &  1.65  &  13.51$\pm$0.06  & 2MASS \\
  $K_s$      &  2.15  &  13.14$\pm$0.16  & DENIS \\
  $K_s$      &  2.17  &  13.33$\pm$0.06  & 2MASS \\
\hline
\end{tabular}
\end{table}

Note that different measurements have been based on observations taken
at different epochs. However, the fairly constant $B$ magnitude obtained in
Goranskij et~al. (\cite{goran}) between 1928--1994 shows that the
progenitor of V838~Mon was not significantly variable.

As can be seen from Table~\ref{mag}, for four photometric bands we have two
independent measurements. In the case of the $B$ and $J$ magnitudes the agreement
is good. The values in the $I$ and $K$ bands are discrepant by
$\sim$0.2~magnitude. As it is difficult to judge which result is more reliable,
for futher analysis we have adopted mean values in the bands for which two
measurements have been available.

An analysis of the progenitor has to take into account the B-type companion
discovered by Munari et~al. (\cite{mundes}). It accounts for about half the brightness
of the progenitor. It seems most reasonable to assume that V838~Mon
and its B-type companion form a binary system.
The main argument obviously comes from the observed positions.
From the instrumental crosses of stars seen on the HST images
taken in September--December 2002
(http://hubblesite.org/newscenter/archive/2003/10/,
see also Bond et~al. \cite{bond}) one can
deduce that the central object in the B images (dominated by the B-type
companion) very well coincides with the central object in the I image
(dominated by V838~Mon itself) and that both stars cannot be separated by
more than $\sim 0\farcs 1$.
In the HST field ($83\arcsec \times 83\arcsec$)
there are $\sim$10 field stars of similar brightness as V838~Mon
before outburst and the B-type companion.
In this case the probability that due to a random coincidence one of
these stars is separated by $\la 0\farcs 1$ from
V838 Mon is $\la 10^{-4}$. 
Next, as discussed below, the binary hypothesis leads to
a consistent interpretation of the observational data of the progenitor.
Observational determination of the distance and reddening
to V838~Mon itself and its B-type companion,
summarized and discussed in Tylenda (\cite{tyl05}), give consistent results,
in the sense that
there is no significant difference in the results for both objects.
Therefore in most of our discussion we assume that V838~Mon and its B-type
companion are at the same distance and suffer from the same interstellar
extinction. In some cases, however, we relax this assumption and discuss
the consequences of that.

\begin{figure*}
\centering
  \includegraphics[width=8.5cm]{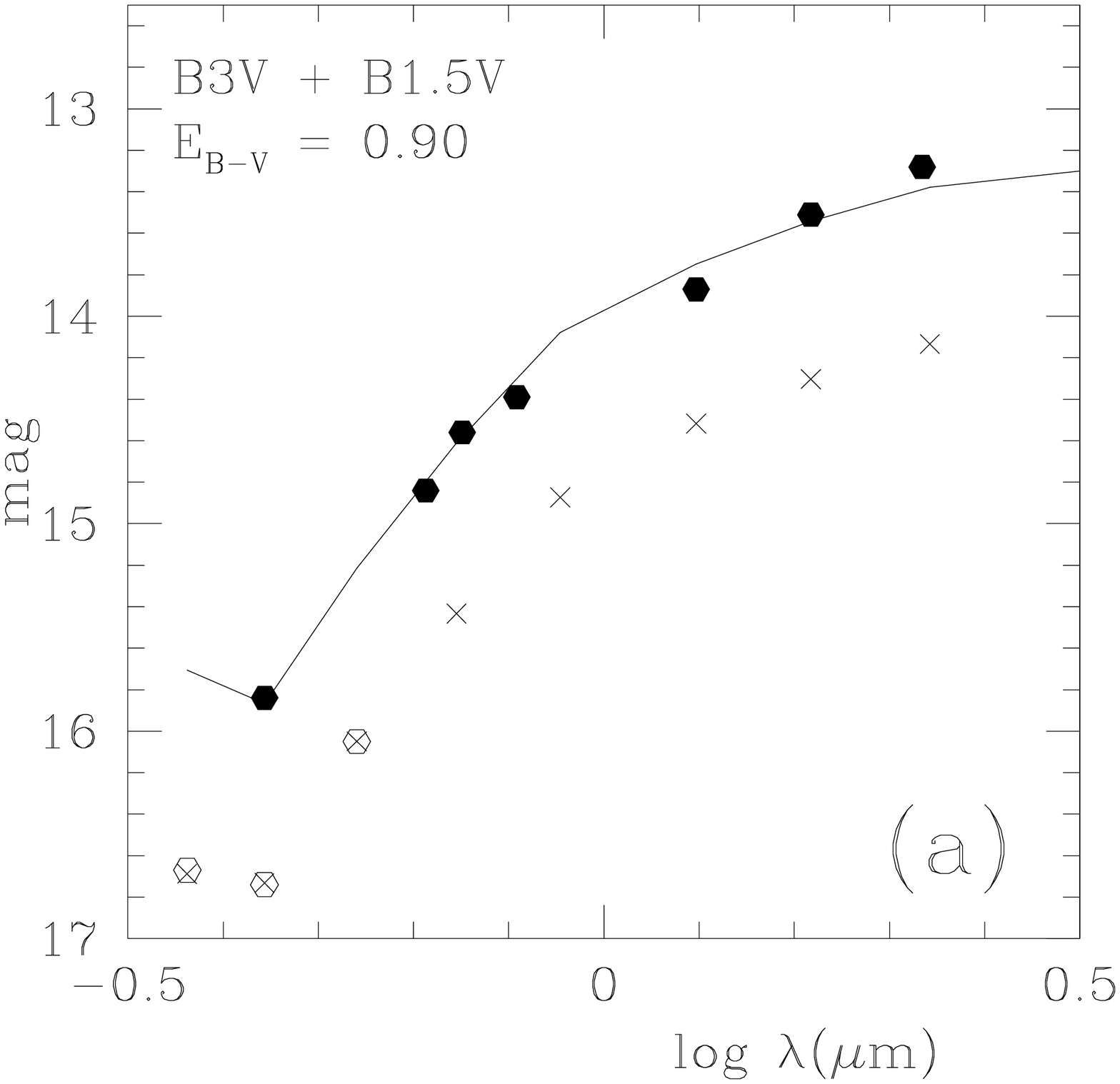}
  \includegraphics[width=8.5cm]{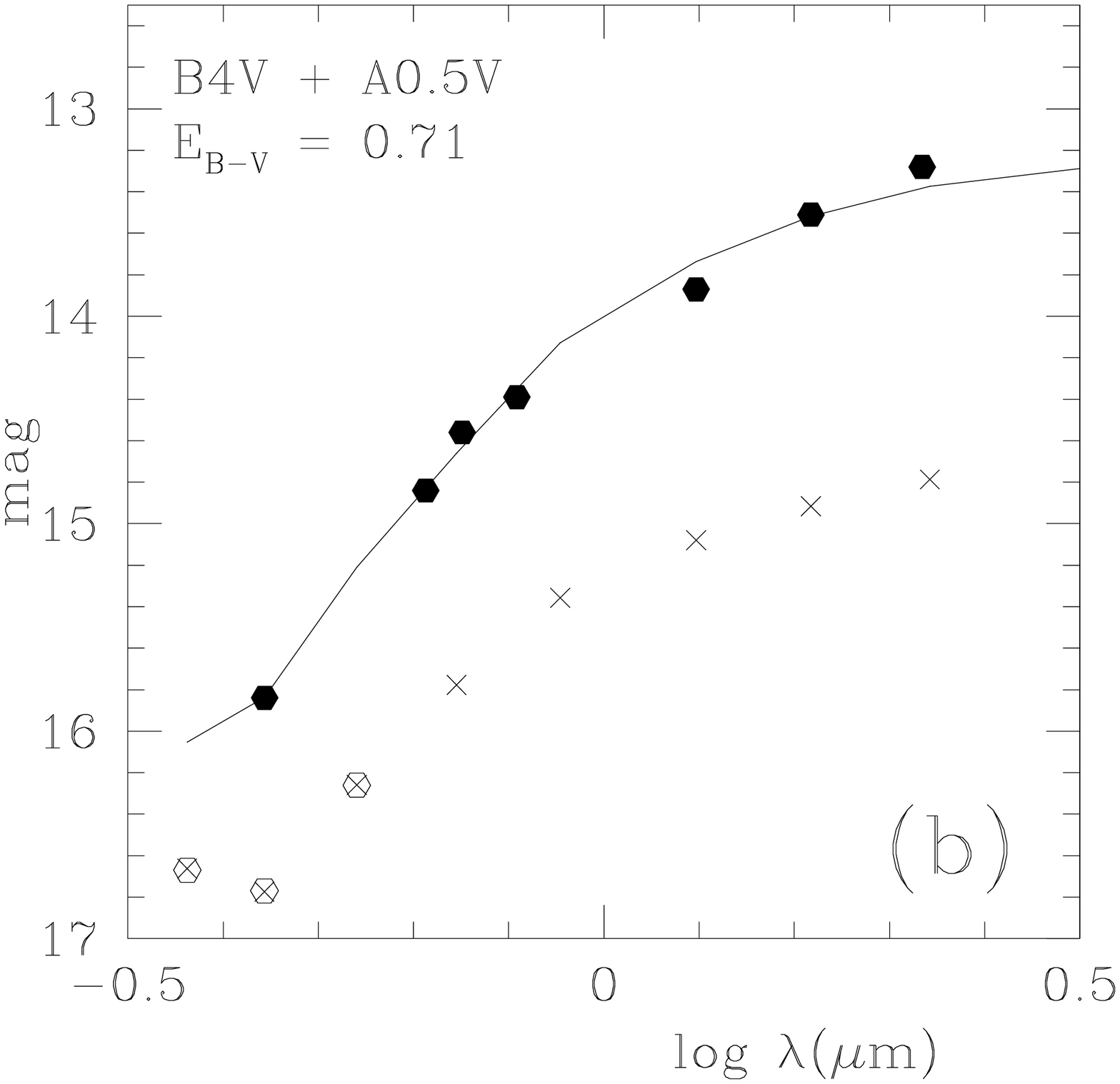}
  \caption{Spectrophotometry of the progenitor of V838~Mon.
Full symbols -- observed magnitudes from Table~\ref{mag}.
Part {\bf (a)} (left panel): $UBV$ magnitudes of the B-type companion
(open symbols) and $E_{B-V} = 0.90$ adopted from Munari et~al. (\cite{mundes}).
Crosses -- standard B3~V star, full curve -- standard B3~V and B1.5~V stars
co-added. Part {\bf (b)} (right panel): $UBV$ magnitudes of the B-type companion
(open symbols) and $E_{B-V} = 0.71$ derived from Crause et~al. (\cite{lisa04})
(see text). Crosses -- standard B4~V star, full curve -- standard B4~V
and A0.5~V stars co-added. General note: all the model spectra (crosses
and full curves) have been reddened with the respective values of $E_{B-V}$.
}
  \label{prog}
\end{figure*}

As mentioned above, from spectroscopy Munari et~al. (\cite{mundes}) have identified
the hot companion of V838~Mon
as a typical B3~V star. Indeed their photometric results obtained in September and
October~2002, i.e. $V =16.05$, $B-V = 0.68$ and $U-B = -0.06$ (Munari et~al.
\cite{munhen}), can be well
reconciled with the standard B3~V colours (see Schmidt-Kaler \cite{sk}) provided
that the object is reddened with $E_{B-V} = 0.90$ (see open symbols and crosses
in Fig.~\ref{prog}{\bf a}).

The brightness of the B-type companion can also be deduced from the photometric
results of Crause et~al. (\cite{lisa04}). Between September~2002 and January~2003
(AJD 529--668 in Table~2 of Crause et~al.) V838~Mon was
practically constant in $UBV$ while steadily brightening in $R$ and $I$. This
behaviour of the object was also noted by 
Munari et~al. (\cite{mundes}, \cite{munhen}) and indicates, in
accord with their spectroscopic results, that the $UBV$ magnitudes were dominated
by the B-type companion during this time period. From the most reliable results of
Crause et~al., i.e. those for dates not marked with an asterisk in their Table~2, obtained
between AJD~529--668 one derives (mean value $\pm$ standard deviation)
$V = 16.26 \pm 0.02$, $B-V = 0.51 \pm 0.02$, and $U-B \simeq -0.1$. Thus the object
is by $\sim$0.2~mag. fainter in $V$ than in Munari et~al., while the $B-V$ value
if interpreted with the B3~V standard gives $E_{B-V} = 0.72$. In this case however
the object seems to be not blue enough in $U-B$. A better agreement with
the above results derived from Crause et~al. is obtained for a B4~V standard
and $E_{B-V} = 0.71$ (see open symbols and crosses
in Fig.~\ref{prog}{\bf b}).

From the beginning of October~2002 V838~Mon has also been
measured by Goranskij et~al. (\cite{goran}).
From their results obtained in October--December~2002 one
derives $V = 16.17 \pm 0.05$ and $B-V = 0.64 \pm 0.13$ (no $U$ measurements have been
done during this time period). Thus the $V$ magnitude is in between the values of
Munari et~al. and Crause et~al., while the $B-V$ value is closer to that of
Munari et~al. and implies $E_{B-V} = 0.84 \pm 0.14$.
A large scatter in the $B$ measurements of Goranskij et~al.
should however be noted.

Later on in this section we consider two cases depending on whether
the photometric data for the B-type companion are adopted from 
Munari et~al. (\cite{munhen})
or from Crause et al. ({\cite{lisa04}). The differences in the magnitudes between
these two references are extreme (the data from Goranskij et al. \cite{goran}
are in between them) so these two cases allow us to see how the results of our analysis
depend on uncertainties in the photometry of the B-type companion.

Figure~\ref{prog} presents our interpretation of the available photometric
data done assuming that both V838~Mon and the B-type companion have
the same reddening. In the discussion we also assume that both components are
at the same distance, namely that they form a binary system.
In both parts of the figure full symbols display
the observed magnitudes from Table~\ref{mag}.
The best fits, shown with full curves, have been made using the least square method
and the intrinsic photometric colours
for the main sequence stars
taken from Schmidt-Kaler (\cite{sk}), Johnson (\cite{johnson}),
Koornneef (\cite{koor}) and Bessell \& Brett (\cite{bess}) (for more details
on the fitting procedure see Tylenda \cite{tyl05}).

In part {\bf (a)} open symbols show the $UBV$ photometry of
the B-type companion taken from Munari et~al. (\cite{munhen}) fitted with
a standard B3~V star shown with crosses.
The full curve presents the best fit to the full points obtained with
a standard B1.5~V star
added to the B3~V companion. Both spectral components have been reddened with
$E_{B-V} = 0.9$. The ratio of the luminosity of the B1.5 component to that of B3
is 1.9. This ratio is somewhat too low for the B1.5~V and B3~V stars
but given the uncertainties in the observational data we can conclude
from Fig.~\ref{prog}{\bf (a)} as follows. The progenitor of V838~Mon was a binary
system consisting of two early B main sequence stars. V838~Mon itself was
probably somewhat brighter, hotter and more massive than its companion.
The system is young, i.e. $\la 2 \times 10^7$~yrs (main sequence lifetime
of a 9~$M_{\sun}$ star, typical for B2~V). Note that it is excluded that
V838~Mon was an evolved B1.5 star as then it would have been significantly more
luminous than the B3 main sequence companion.

Fig.~\ref{prog}{\bf (b)} adopts the parameters of the B-type companion derived
from the photometry of Crause et~al. (\cite{lisa04}), i.e. a B4~V star reddened
with $E_{B-V} = 0.71$. In this case in order to reproduce the observational
data for the progenitor an A0.5~V standard star has been added to the B4~V companion.
The luminosity ratio of the A component to the B one is 0.43. This is much larger
than the ratio for the main sequence of the same types which, according to
Schmidt-Kaler (\cite{sk}), is $\sim 0.02$.\footnote{From this result one may consider
that V838~Mon was an A-type main sequence
star prior to eruption, thus rejecting the binarity hypothesis.
However in the case of the A0.5 main sequence
star it would be at a distance of 3.0~kpc which is too low given the distance
estimates from the light echo.}
Also the possibility that the A component was an evolved star, e.g.
a giant evolving towards the red giant branch (RGB), the AGB, or a post-AGB
star, can be ruled out.
In this case it would be expected to have been initially (while being on
the main sequence) more massive and thus, at present, significantly more luminous
than the B4 main sequence companion. Therefore the only possibility within
the binary hypothesis is that the A0.5 star is in the pre-main-sequence phase.
The system would thus be very young.
Judging from the luminosity of the A-type component, $\sim 550\ L_{\sun}$ if
1300~$L_{\sun}$ is assumed for the B4~V component,
its mass would be $\sim 5~M_{\sun}$ and the age of the system
would be of $\sim 3 \times 10^5$~yrs (Iben \cite{iben}).

In summary, although the uncertainties in the observational data for the
progenitor and for the B-type companion do not allow us to unambiguously
identify the nature of V838~Mon the above discussion allows us to put rather
narrow constrains if the most probable hypothesis of binarity is adopted.
In this case V838~Mon is a system
consisting of two intermediate mass stars.
V838~Mon itself certainly was not an evolved star, e.g. RGB, AGB, post-AGB.
It is either slightly more massive than its B-type companion, i.e. 8--10~$M_{\sun}$,
and was on the main sequence
prior to eruption, or is somewhat less massive, $\sim 5\ M_{\sun}$, being in the
pre-main-sequence phase. The system is young, with the age estimated between
$3 \times 10^5$ and $2 \times 10^7$~yrs.
The abundances in V838~Mon obtained by Kipper et~al. (\cite{kipper}) are
reminiscent of those in the so-called HAEBE stars (e.g. Acke \& Waelkens \cite{waelkens})
which are believed to be more massive analogues of the T~Tauri stars.
Therefore it is likely that the V838~Mon system is
still partly embedded
in the interstellar complex from which it has been formed. Indeed,
as discussed in Sect.~\ref{ism}, near the position of V838~Mon there are
several star-forming regions with radial velocities close to that of V838~Mon
and its B-type companion.
This also fits well
the conclusion of Tylenda (\cite{tyl}) and Sect.~\ref{echo} that the circumstellar
dust producing the light echo of V838~Mon is most probably of interstellar
origin.

Munari et~al. (\cite{munhen}) have made an analysis of
the photometric data for the V838~Mon progenitor similar to ours. 
Their conclusion is qualitatively similar to ours in the sense that
the progenitor was an early-type star.
However, contrary to our main-sequence or pre-main-sequence hypothesis,
Munari et~al. conclude
that the V838~Mon outburst was that of an evolved star of initial mass
of $\sim 65\ M_{\sun}$, at present in a region occupied by Wolf-Rayet stars in the
HR diagram and having $T_\mathrm{eff} \simeq 50\,000\ $K.
However, in a case like this
we should see a bright HII region surrounding V838~Mon. The observed light
echo (discussed in Sect.~\ref{echo}) shows that there is a lot of 
diffuse matter extending from $\sim 0.1\ $pc up to at least $\sim 4\ $pc.
The 50\,000~K star of Munari et~al. (\cite{munhen}), assuming $E_{B-V} = 0.9$ and
a distance of 8~kpc, would have a luminosity of $\sim 3 \times 10^4\ L_{\sun}$.
Using model results of Stasi\'nska (\cite{stas}), for abundances depleted
by a factor of 2 relative to standard values (V838~Mon lies at the outskirts of the
Galactic disc), we can estimate that a star like
this would be able to ionize the surrounding matter up to $R_\mathrm{s} \simeq 8$~pc
if its density is $n_\mathrm{H} = 10$~H~atoms~cm$^{-3}$ ($R_\mathrm{s}$
scales as $n_\mathrm{H}^{-2/3}$). The emission line spectrum would
be dominated by [OIII] and Balmer lines 
([OIII]$\lambda 5007$\AA /H$\beta \simeq 8$), while 
the H$\beta$ luminosity would be $\sim 230\ L_{\sun}$. For an observer (at 8~kpc
and $E_{B-V} = 0.9$) it would look like a nebula with a diameter of $\sim 7\arcmin$
and an H$\beta$ flux of 
$\sim 6 \times 10^{-12}$~erg~s$^{-1}$~cm$^{-2}$. The resultant H$\beta$ surface
brightness of $\sim 4.5 \times 10^{-17}$~erg~s$^{-1}$~cm$^{-2}$~arcsec$^{-2}$
is typical for many extended planetary nebulae, e.g. those in the Abell
(\cite{abell}) catalogue (observed H$\beta$ fluxes and nebular diameters 
can be found in Acker et~al. \cite{acker}).
Thus the nebula would be rather easy to discern observationally,
especially that in H$\alpha$ and [OIII]$\lambda 5007$\AA~it would 
$\sim 8$ times brighter than in H$\beta$.
Yet no emission-line nebula have been discovered around the position
of V838~Mon (Orio et~al. \cite{orio02}, Munari et~al. \cite{munari}).
Thus the idea of Munari et~al. (\cite{munhen}) that V838~Mon prior outburst 
could have been as hot as 50\,000~K is not consistent with the observations.
From the observed lack of any significant emission nebula around V838~Mon
we can conclude that before the outburst the star was cooler than $\sim 30\,000$~K,
i.e. of a spectral type not earlier than B0. Munari
et~al. (\cite{munhen}) also consider that the progenitor
could have had $T_\mathrm{eff} \simeq 25\,000\ $K (although they argue that this
is not likely). This  solution is practically the same as our case
of a B1.5 star in Fig.~\ref{prog}{\bf (a)} 
which we interpret as an early B-type main sequence star.

As discussed above it is evident that V838~Mon was not a typical red giant
nor an AGB star prior to eruption
if V838~Mon and its B-type companion form a binary system.
The only way to reconcile the RGB or AGB hypothesis is to assume that
the B-type companion has nothing to do with V838~Mon and that the coincidence of
the two objects in the sky is purely accidental. Then one may assume that V838~Mon
is less reddened than the B-type companion and a cooler star can be fitted to
the observations. Let us consider that the B-type companion has the parameters
derived from the observations of Crause et~al. (\cite{lisa04}), i.e. B4~V reddened
with $E_{B-V} = 0.71$, as then the fits give later spectral types for V838~Mon
than if the results of Munari et~al. (\cite{mundes}) were adopted.
Let us also use the standard supergiant spectra (intrinsic colours taken from the
same references as the main sequence ones) to model the contribution
from V838~Mon. This is more relevant with the RGB/AGB hypothesis and also results in
later spectral types from the fits than the main sequence spectra.
$E_{B-V} = 0.5$ seems to be a lower limit for the extinction towards V838~Mon
(see discussion of different observational determination in Tylenda \cite{tyl05}).
Assuming this value, the best fit to the observations is obtained for the spectral
type F1 (effective temperature $\sim 7500$~K). If, in spite of observational
determination, the extinction is pushed to its limit, i.e. $E_{B-V} = 0.0$ is
assumed, the fit gives G7 (effective temperature $\sim 4700$~K). Thus there is
no way to reconcile an M-type star with the observational data.
This conclusion is obvious if one realises
that the $B-R$ colour for an unreddened M-type star is $\ge 2.7$ while that of
the V838~Mon progenitor was $\simeq 1.3$.

From the above results we can firmly conclude that V838~Mon was not an
AGB star. If one still does not want to leave the AGB hypothesis and argues that it
best explains the existence of the circumstellar matter seen in the echo and infrared
images, then the only way to reconcile it with the photometric data is to say that
V838~Mon had quite recently left the AGB and prior to eruption was in the post-AGB phase.
However in this case its luminosity would be close to $5 \times 10^3 L_{\sun}$ and its
distance would have to be $\sim 55$~kpc in the case of the F1 type and $E_{B-V} = 0.5$ or
even $\sim 90$~kpc if one prefers G7 and $E_{B-V} = 0.0$. Thus a typical spectral types
of the post-AGB stars (F--G) would require rather unacceptable conditions, i.e.
a very low reddening and
a very large distance putting V838~Mon in the extreme outskirts of the Galaxy.
For the observationally acceptable range of the extinction,
i.e. $E_{B-V}$ between 0.7 and 1.0 (Tylenda \cite{tyl05}), V838~Mon would have been
of the B9--B0 type and, assuming the typical post-AGB luminosity as above, its distance
would be between 7 and 30~kpc. Thus the only possibility within the AGB
-- post-AGB hypothesis, which is not excluded by
the photometric data, is that V838~Mon was a B-type post-AGB star.
However, as discussed above, the binarity with the B-type main
sequence component is, in this case, excluded in spite of similar estimated distance
ranges (7.5--12.5~kpc for the B-type companion, see Tylenda \cite{tyl05}), close
values of interstellar extinction and the same positions in the sky of both objects.
If one adds that the duration of the phase when the post-AGB star can be classified
as B-type is typically $10^3$~years it is clear that this solution is extremely
improbable.

Finally let us discuss the giant hypothesis. As discussed above it is certain that
V838~Mon was not a typical RGB star, i.e. of K--M spectral type, prior to eruption.
The latest acceptable spectral type, obtained assuming the lower limit of
$E_{B-V} =  0.5$, is F0--A5. At the lower limit of the distance of 5~kpc
(Tylenda \cite{tyl05}) the star would have a luminosity of 40--50~$L_{\sun}$
which is more or less consistent with the standard giant luminosities for
these spectral types (Schmidt-Kaler \cite{sk}). For the more probable reddening,
i.e. $E_{B-V} = 0.7 - 1.0$, we have to move to the B types and correspondingly larger
luminosities and distances. Thus the hypothesis that V838~Mon was a giant before
its eruption is not excluded provided that it was an early (A--B) type giant.
The binarity with the B-type companion is excluded in this case (as discussed above).
The star would be quite massive, $\ga 2.5 M_{\sun}$, not far evolved
from the main sequence and thus be in a fast evolutionary phase (time scale
$\la 5 \times 10^6$~years). The object would be quite rare in the stellar
population although not as rare as the B-type post-AGB case considered above.
The discussed case would however have difficulties in explaining the origin
of the circumstellar matter seen in the light echo. The wind from an early type
giant would not be enough. On the other hand, the giant, being significantly
older (most probably at least as old as $10^8$ years) than the B-type binary system
considered above, would have little chance to still reside in a dense interstellar
cloud.

\section{Circumstellar and interstellar enviroment  \label{enviroment}}

\subsection{The light echo  \label{echo}}

The phenomenon of light echo observed in V838~Mon during and after the outburst
suggests that there is much dusty matter in the vicinity
of the object. Since the light echo works as a sort of scanner, an analysis
of the light echo images in different epochs should provide detailed information
on the dust distribution near the object which would be important for constraining
the nature of the object. Unfortunately, in spite of numerous images
obtained at different observatories (including HST) no elaborate study of the
light echo has been done as yet. So far the most detailed, but still very
simple, analysis has been done by Tylenda (\cite{tyl}) on five
echo images obtained with HST between 30~April and 17~December~2002.
His conclusion is that the dust distribution does not show any signs of
spherical symmetry and that dust is likely to be of interstellar origin
rather than due to past mass loss from V838~Mon.

We have extended the analysis of Tylenda (\cite{tyl})
using two recent echo
images obtained on 21~Oct.~2003 at the USNO
(http://www.ghg.net/akelly/v838lar3.jpg) and on 8~Feb.~2004 with the HST
(http://hubblesite.org/newscenter/newsdesk/archive/releases/2004/10/).
On these images we have measured the outer rim of the light echo.
Then a least square fit of a circle to
the measurements has been done in the same way as in
Tylenda (\cite{tyl}). The results of the fits are given in the last two lines
of Table~\ref{echo_tab}. The first five lines in the table repeat the results
from Table~1 of Tylenda (\cite{tyl}) as the uncertainties there have been
slightly overestimated. The first column of Table~\ref{echo_tab}
shows the time of observations, $t_\mathrm{obs}$, given in days since
1~January~2002.
The radius of the echo, $\theta$, and its uncertainty
are given in the second column. The next
two columns show the ($x,y$) position of the centre of the fitted circle relative
to the central star. Note that $x$ points to west while $y$ is to north.
The last column gives the angular distance of the echo centre
from the central star.
All the results are in arcsec.
Following Tylenda (\cite{tyl}) we adopt in our analysis that the zero age
of the echo is $t_0 = 34$~days (since 1~Jan.~2002).

\begin{table}
\centering
\caption{Results of fitting a circle to the outer edge of the light echo
         of V838~Mon. Time of observations, $t_\mathrm{obs}$, is in days
         since 1~January~2002. Results are in arcsec.}
\label{echo_tab}
\begin{tabular}{c c c c c}
\hline
  $t_\mathrm{obs}$ & $\theta$ & $x_\mathrm{c}$ & $y_\mathrm{c}$ &
  $\theta_\mathrm{c}$\\
\hline
  120.0 & 18.55$\pm$0.62 & $-$0.83$\pm$0.90 & 0.46$\pm$0.86 & 0.95$\pm$0.88 \\
  140.0 & 20.89$\pm$0.61 & $-$1.12$\pm$0.88 & 0.50$\pm$0.84 & 1.23$\pm$0.86 \\
  245.0 & 30.32$\pm$0.83 & $-$2.27$\pm$1.21 & 0.37$\pm$1.15 & 2.30$\pm$1.18 \\
  301.0 & 33.68$\pm$0.84 & $-$2.36$\pm$1.23 & 1.11$\pm$1.17 & 2.60$\pm$1.20 \\
  351.0 & 36.66$\pm$0.90 & $-$2.61$\pm$1.31 & 1.58$\pm$1.23 & 3.05$\pm$1.27 \\
  659.0 & 52.10$\pm$1.15 & $-$6.05$\pm$1.65 & 3.12$\pm$1.61 & 6.80$\pm$1.63 \\
  769.0 & 56.67$\pm$1.19 & $-$5.91$\pm$1.66 & 5.37$\pm$1.71 & 7.99$\pm$1.69 \\
\hline
\end{tabular}
\end{table}

\begin{figure*}
\centering
  \includegraphics[width=8.5cm]{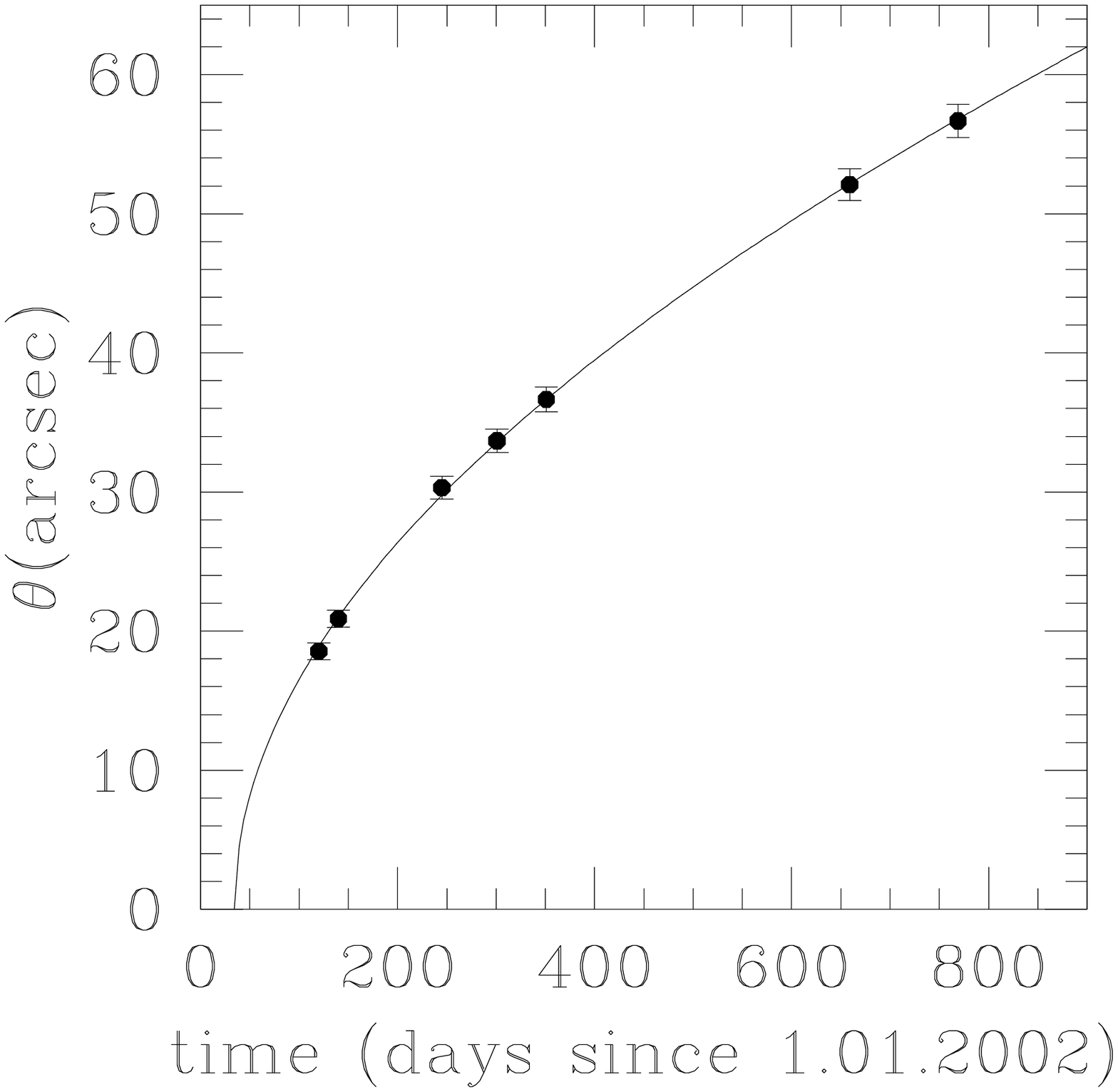}
  \includegraphics[width=8.5cm]{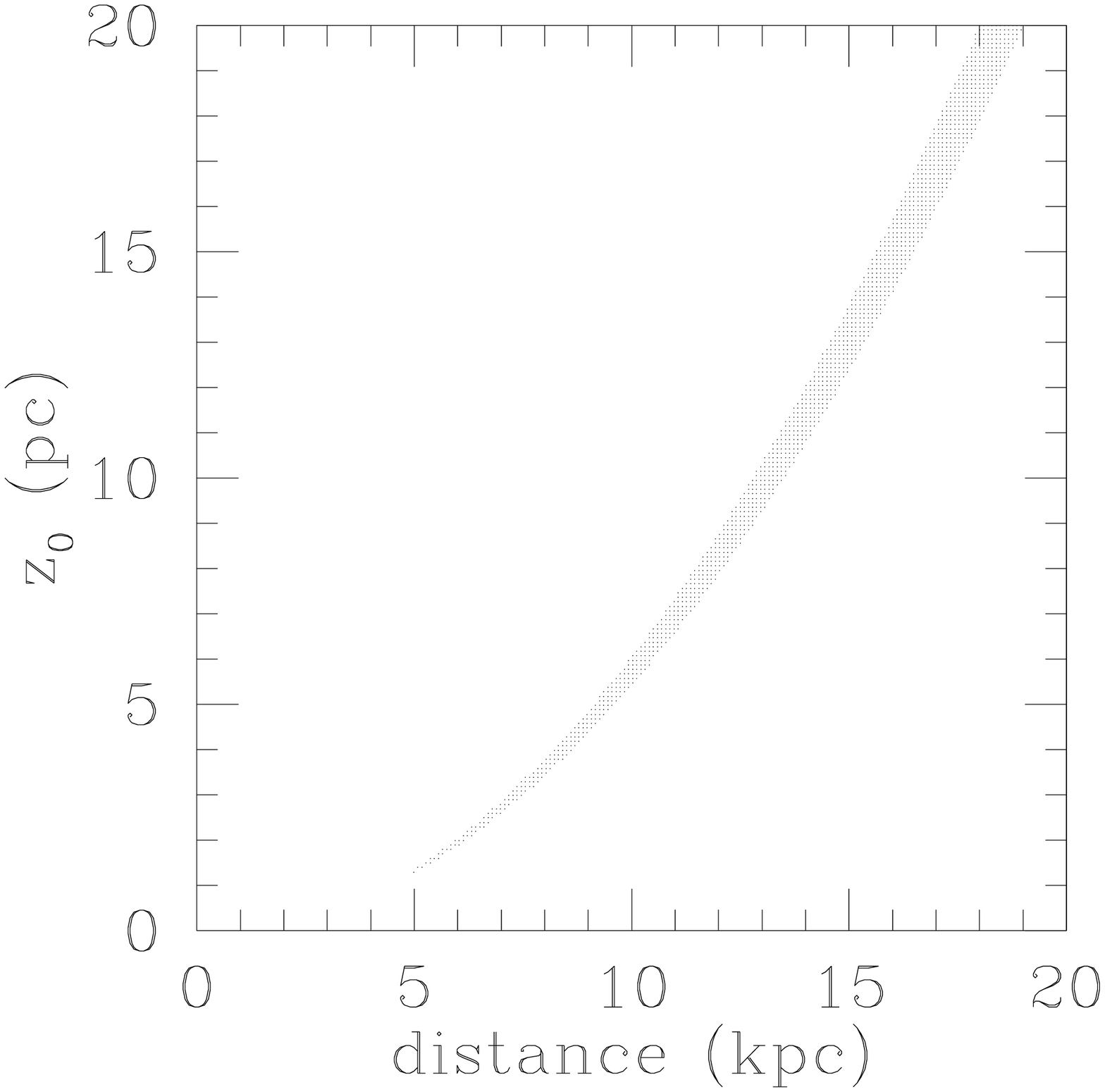}
  \caption{\emph{Left panel:}
          The best fit of a plane model to the observed evolution
           of the light echo radius. Symbols with error bars --
           the values and uncertainties of $\theta$ from column (2) of
           Table~\ref{echo_tab}.
           Full curve -- model predictions for $d = 11.1$~kpc and $z_0 = 7.1$~pc
           (see text).
           \emph{Right pannel:}
           The 95\% confidence region (hatched) of
           the plane model fitted to the observed light
           echo expansion (symbols in the left panel).}
  \label{dz_fig}
\end{figure*}

Filled symbols in the left panel of Fig.~\ref{dz_fig} display
the evolution of the echo radius,
$\theta$, with time. The full curve shows the best fit to the data
of a plane slab model,
i.e. of Eq.~(17) in Tylenda (\cite{tyl}), obtained
for $d \simeq 11.1$~kpc and $z_0 \simeq 7.1$~pc, where $d$ is the distance
between the light source and the observer while $z_0$ is the distance of
the dust slab from the source.
However, similarly to
Tylenda (\cite{tyl}), the $\chi^2$ minimum of the fit is
quite shallow and extended along $z_0 \sim d^2$. The right panel of
Fig.~\ref{dz_fig} shows the 95\% confidence region of the fit.
From this figure one can conclude that
the distance to V838~Mon is $\ga 5$~kpc. Recently, Crause et~al. (\cite{lisa04})
have analysed the light echo evolution from their observations done
at the SAAO. Their results obtained for the sheet model are well within the
hatched region in the right panel of Fig.~\ref{dz_fig}.

\begin{figure*}
\centering
  \includegraphics[width=8.5cm]{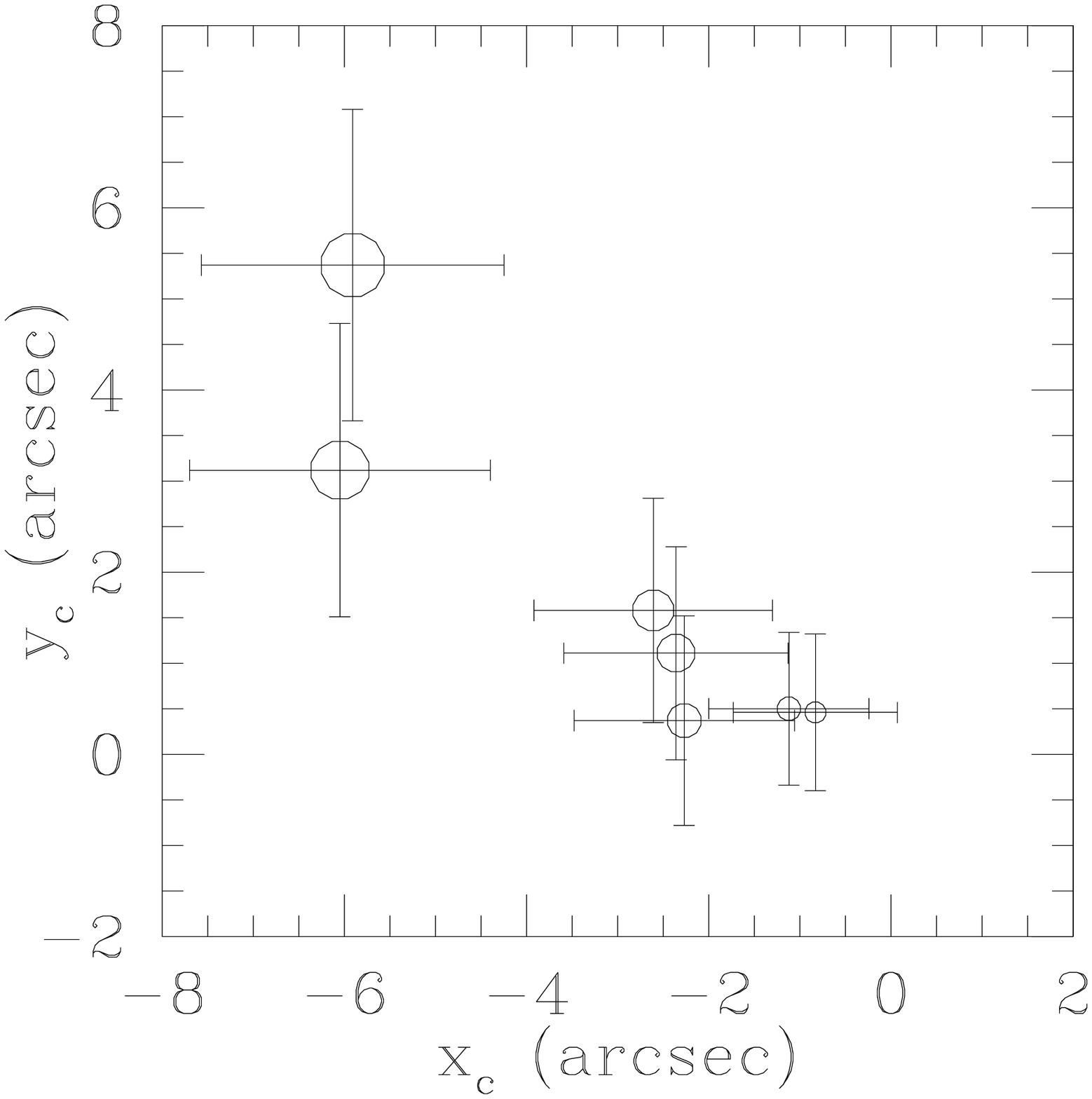}
  \includegraphics[width=8.5cm]{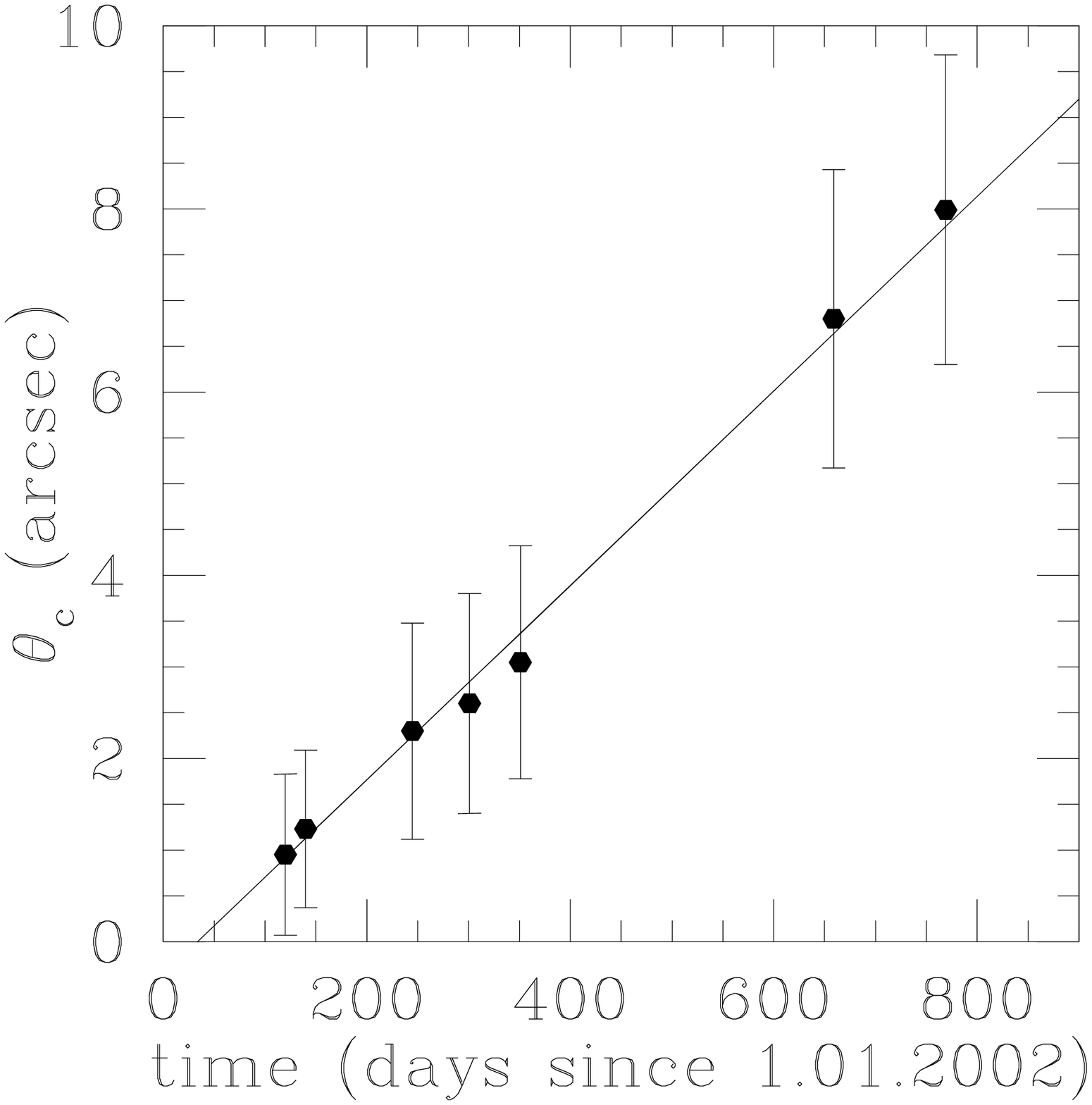}
  \caption{Migration of the echo centre from the central star.
          \emph{Left panel:} ($x$,$y$) positions of the echo centre
          from columns (3) and (4) of Table~\ref{echo_tab}. The size of
          the symbols is proportional to the echo radius given in
          column (2) of Table~\ref{echo_tab}. Note that
          the $x$ and $y$ axes point to west and north, respectively.
          The central star is at ($x=0, y=0$).
          \emph{Right panel:} evolution of the distance of the
          echo centre from the central star, $\theta_\mathrm{c}$,
          with time. Symbols -- the data from column (5) in Table~\ref{echo_tab}.
          Full line -- a linear fit to the data as given in Eq.~(\ref{cent_eq}).}
  \label{cent_fig}
\end{figure*}

As can be seen from Table~\ref{echo_tab}, the centre of the light echo
has been migrating from the central object. This migration, displayed
in Fig.~\ref{cent_fig}, has kept the same pattern for $\sim 2$~years.
The following two conclusions can be drawn from
Fig.~\ref{cent_fig}.
First, since the appearance of the light echo
its center has been moving away from the
central object in roughly the same (north-east) direction,
as can be seen from the left panel
of Fig.~\ref{cent_fig}. Second, the distance of
the echo centre from the central star has been increasing linearly with time,
as shown in the right panel of Fig.~\ref{cent_fig}.

As discussed in Tylenda (\cite{tyl}),
the fact that the light echo has an outer edge
means that the dusty medium producing the echo has a boundary in front of
the central star. However,
the fact that the echo edge is \emph{not} centered on
the star and that the distance of the echo centre from the star \emph{does}
increase with time shows that this dust boundary is \emph{not}
spherically symmetric with respect to the central object.

Following the theory of the light echo, as e.g. summarized in Tylenda (\cite{tyl}),
the only reasonable interpretation of the observed evolution of the
outer edge of the light echo is that the dust boundary in
front of V838~Mon is more or less in the form of a plane inclined to the line
of sight.
A linear fit to the observed evolution of the distance of the echo centre
from the central object, i.e. last column in Table~\ref{echo_tab} or
symbols in the right panel of Fig.~\ref{cent_fig},
gives the relation
\begin{equation}
  \label{cent_eq}
  \theta_\mathrm{c} = (0\farcs0106 \pm 0\farcs0030)\ (\frac{t}{1\,\mbox{day}})
\end{equation}
where $t = t_\mathrm{obs} - t_0$ is the time since the zero age of the echo.
This relation is shown by a full line in the right panel of Fig.~\ref{cent_fig}.
Assuming a distance of 8~kpc
Eq.~(\ref{cent_eq}), together with Eq.~(7) of Tylenda (\cite{tyl}), implies
that the normal to the dust surface is inclined to
the line of sight at an angle of $\sim26\degr$.
This surface is at a distance of $\sim 3.5$~pc from V838~Mon
(along the line of sight) and its portion so far (i.e. till Feb.~2004)
illuminated by the light echo has dimensions of $\sim 4.4 \times 4.9$~pc.
It is difficult to imagine that such a large flat surface of the dusty medium
could have been produced by mass loss from V838~Mon itself. Instead, it suggests that
the dusty medium in the vicinity of V838~Mon is much more extended than
the illuminated part, thus most probably being of interstellar origin.

In April~2002, when V838~Mon faded in the optical, the light echo started developing
an asymmetric hole in the centre. As analyzed in Tylenda (\cite{tyl})
this clearly shows that there is
a dust-free region around V838~Mon and that this empty region is strongly asymmetric.
The inner edge of the dusty region in the southern directions
is at 0.10--0.15~pc from the central object whereas in the opposite directions
it is at least 10 times further away.
It can be noted that the IRAS PSC source 07015$-$0346 coincides
with the position of V838~Mon within
a position elipse of $48\arcsec \times 10\arcsec$
(see also Kimeswenger et~al. \cite{kimes}). The
measured fluxes at 100~$\mu$m and 60~$\mu$m are 4.6~Jy and 1.4~Jy, respectively,
while at 25~$\mu$m and 12~$\mu$m the catalogue gives only an upper limit of
0.25~Jy. When fitted with a simple dust emission model, i.e. emissivity
proportional to $\lambda^{-1}B_{\lambda}(T_\mathrm{d})$,
the IRAS fluxes give a dust temperature, $T_\mathrm{d}$, of
$\sim 30$~K. For a central source of $10^4\ L_{\sun}$ (two B3~V stars, see
Sect.~\ref{phot}) this value of dust temperature is reached at $\sim 0.2$~pc
(see e.g. Eq.~7.56 in Olofsson \cite{olof}).
Thus the IRAS fluxes can be consistently interpreted as due to inner parts
of the dusty region inferred from the light echo analysis. In particular,
they give evidence that there is no significant amount of dust at distances
$\la 0.1$~pc, thus confirming the existence of the central dust-free region.

The strongly asymmetric central dust-free region would be very difficult
to understand if the echoing dust were produced by a past mass loss from V838~Mon.
The hole would imply that mass loss stopped a certain time ago, e.g. $10^4$~yrs
for the 0.10--0.15~pc inner dust rim if a wind velocity of
$10~{\rm km}~{\rm s}^{-1}$ is assumed.
However the hole asymmetry
would imply that in the opposite direction either mass loss stopped 10 times earlier
or the wind velocity was 10 times higher. Neither of these two possibilities seems
to be likely.

Instead, as discussed in Tylenda (\cite{tyl}), the asymmetric hole
is easy to understand if the echoing dust is of interstellar origin. Then it is
natural to suppose that V838~Mon is moving against the ISM.
If possessing
a fast wind it would create a hole largely asymmetric along the direction of
the movement. Indeed, the structure of the inner edge of dust in Fig.~5 of
Tylenda (\cite{tyl}) well resembles stellar wind bow shocks investigated in
e.g. Van~Buren \& McCray (\cite{vanbur}) and Wilkin (\cite{wilk}).
The nearest rim in the southern directions, being
at 0.10--0.15~pc from the central star, would correspond to a region where
the stellar wind collides head-on with the ambient medium.

Let us assume that a star losing mass at a rate,
$\dot{M}_\mathrm{w}$, and a velocity, $v_\mathrm{w}$, is moving in the ISM
of number density, $n_0$, with a relative velocity of $v_\ast$.
A swept-up shell is created in the form of a bow shock and in the up-stream direction
this takes place where the wind ram pressure is comparable
to that of the ambient medium (see e.g.
Van~Buren \& McCray \cite{vanbur}, Wilkin \cite{wilk}), i.e.
\begin{equation}
  \label{press_eq}
  v_\ast^2\ n_0 \simeq v_\mathrm{w}^2\ n_\mathrm{w}
\end{equation}
where $n_\mathrm{w}$ is the number density in the wind and is related to the mass
loss rate in a standard way
\begin{equation}
 \label{rate_eq}
 \dot{M}_\mathrm{w} = 4 \pi r^2\ v_\mathrm{w}\ \mu m_\mathrm{H}\ n_\mathrm{w}
\end{equation}
where $m_\mathrm{H}$ is the H atom mass while $\mu$ is the mean molecular
weight in units of $m_\mathrm{H}$. Then Eq.~(\ref{press_eq}) yields
a standoff distance, $r_0$, from the star, i.e.
\begin{eqnarray}
\lefteqn{ r_0 \simeq 0.15 \left( \frac{\dot{M}_\mathrm{w}}{10^{-9}\ 
     M_{\sun}\ \mathrm{yr}^{-1}} \right)^{1/2}
     \left( \frac{v_\mathrm{w}}{1000 ~\mathrm{km} ~\mathrm{s}^{-1}} \right)^{1/2} }
\nonumber\\
& & {} ~~~\left( \frac{v_\ast}{10 ~\mathrm{km} ~\mathrm{s}^{-1}} \right)^{-1}
     \left( \frac{n_0}{1 ~\mathrm{cm}^{-3}} \right)^{-1/2} \mathrm{pc}.
\label{r0_eq}
\end{eqnarray}
where $\mu = 1.4$ has been assumed.

A B3 main sequence star has a luminosity of $5.0 \times 10^3\ L_{\sun}$
(Schmidt-Kaler \cite{sk}).
Thus, according to the relation of Howard \& Prinja (\cite{howard}), the expected
mass loss rate would be
$7 \times 10^{-10}\ M_{\sun}\ \mathrm{yr}^{-1}$.
If V838~Mon is a young binary system, as discussed in Sect.~\ref{phot}, its velocity
relative to the ISM should be rather low. Assuming
$v_\ast$ between $3 - 10\ \mathrm{km}\ \mathrm{s}^{-1}$ and the above
mass loss rate, Eq.~(\ref{r0_eq}) yields $n_0 \simeq 1 - 10$~cm$^{-3}$
if $r_0 = 0.10-0.15$~pc. This result is uncertain but it indicates that V838~Mon
is imbedded in a relatively dense ISM. In principle, the
density could be estimated from the echo brightness.
Unfortunately there is no such estimate available, although
the presence of the bright echo
suggests that the density of the ambient medium must be significant.

A fast stellar wind colliding with a circumstellar medium
should produce X-rays. V838~Mon was observed with Chandra a year after the outburst
by Orio et~al. (\cite{orio}). The object was not detected and the upper limit
to the X-ray luminosity is $\sim 0.13\ L_{\sun}$ (for a distance of 8~kpc).
The kinetic power of a wind with parameters as above 
($7 \times 10^{-10}\ M_{\sun}\ \mathrm{yr}^{-1}$ expanding at 
$1000\ \mathrm{km}\ \mathrm{s}^{-1}$) is
$\sim 0.06\ L_{\sun}$, thus it is below the X-ray limit, although only by a factor of 2.
However, the X-ray luminosity from a colliding wind is usually much below the
wind kinetic power (e.g. Soker \& Kastner \cite{sokkas}). A hot bubble
is usually formed and, if the radiative cooling time is long, the shocked gas cools
adiabatically by expansion of the bubble. In our case the cooling time
is estimated to be above $10^8$~years, i.e. much longer that the life time
of a B3 main sequence star. In the case of a bow-like inner edge
of the circimstellar matter, as discussed above, a hot bubble need not be formed. 
Instead the shocked gas may flow along the edge and escape through the open side,
thus expanding, cooling adiabatically and radiating very little X-rays. 

van Loon et al. (2004) have also analyzed
the expansion of the light echo.
Their analysis has been based on their measurements of the echo diameter on images
from different sources. However, as also noted in Crause et~al. (\cite{lisa04}),
their diameters are
systematically smaller by factor of 2.5--2.7 than any other measurements available
in the literature (Munari et~al. \cite{munari}, Tylenda \cite{tyl},
Crause et~al. \cite{lisa04}, see also numerous individual measurements
in IAU~Circ. in 2002), including our present results in Table~\ref{echo_tab}.
It is curious that these authors do not note this discrepancy and do not
comment on it.
In any case it can be concluded that the whole analysis of the light echo
made by van~Loon et~al. is questionable as it has been based on wrong data.
In particular, their rather speculative interpretation of the outer edge of
the echo in October~2003 -- February~2004 as being produced by scattering under
right angles does not hold. With the correct values of the echo diameter in these
dates this interpretation would imply a distance of $\sim 2$~kpc (and not 5.5~kpc
as written in van~Loon et al). which is much too low compared with any other
distance estimates from the light echo evolution (Bond et~al. \cite{bond},
Tylenda \cite{tyl}, Crause et~al. \cite{lisa04}).

\subsection{Infrared and CO shells of van~Loon et~al.  \label{shells}}

van Loon et al. (2004, hereafter \cite{lers}) from their analysis of 
the IRAS and MSX images and
the CO maps claim that V838~Mon is surrounded by three shells. The innermost,
seen from the MSX, is highly irregular and has dimensions of $\sim 1 \farcm 5$.
The second one would be the elliptical one referred from the IRAS with dimensions of
$15-20 \arcmin$. The largest one, having a diameter of $\sim 1 \degr$, is suggested
from the CO maps. On this basis \cite{lers} conclude that V838~Mon is a low mass
AGB star experiencing thermal pulses. As we have shown in Sect.~\ref{phot},
the photometric data on the V838~Mon progenitor exclude the AGB hypothesis.
However, even if the latter is not taken into account,
we find severe problems with the results of \cite{lers} and
their interpretation. 

First questions arise when analyzing the innermost structure seen
in the band A (8.3~$\mu$m) of MSX. Unlike the two
outer shells (IRAS and CO) showing an elliptical or circular symmetry
this structure is very distorted and does not show any kind of symmetry
with respect to V838~Mon.
It does not resemble objects involving AGB, like
planetary nebulae, envelopes of AGB and post-AGB stars.
Usually, as in planetary nebulae for example, the ISM
affects external regions so one would expect to see more distortion in
the IRAS and CO shells than in the MSX structure. 

The dimensions of the IRAS and CO shells seem to be too large to be
compatible with the hypothesis that they have been produced by an AGB mass
loss. Adopting a distance to V838~Mon of 8~kpc
(Tylenda \cite{tyl}) the radius of the IRAS shell is $\sim 20$~pc while
that of the CO shell is $\sim 70$~pc.
{{{ (Munari et al. \cite{munhen} argue for 10~kpc as the
most probable distance, which would make radii even
larger thus strengthening our conclusions below.)  }}}
The largest observed AGB dust shells
have radii of 2--3~pc (Speck et~al. \cite{speck}) while the CO shells
are usually well below 1~pc (Olofsson \cite{olof}). 
We can thus conclude that the IRAS and CO shells, if real, are not typical
AGB shells.

Below we show that
it is unlikely that AGB shells could survive and be observed as fairly
symmetric structures at distances of 20--70 pc from the central star.
As a mass-losing star moves relative to the ISM, the shell's segment
in the up-stream direction (the side facing the ISM) is slowed down,
until it is stopped at a time $t_{\rm stop}$, when the leading edge is at a
distance $r_{\rm stop}$ from the star.
Next the up-stream segment of the shell is pushed by
the ISM toward the central star.
At the same time the shell's segment in the down-stream direction
is expanding at a constant, undisturbed rate.
Soker et~al. (\cite{soker}) have derived simple analytical expressions for these
parameters which can be used here.

Let a shell of mass, $M_s$, expand with a velocity $v_{\rm exp}$,
and let $v_\ast$ be the relative velocity of the mass-losing star and
the ISM.
Also let $\rho_0$ be the mass density of the ISM, and $n_0$
the total number density of the ISM. For a distance of
$\sim$150~pc from the galactic plane, we can scale the ISM
density with $\rho_0=10^{-25}\,\mathrm{g}\,\mathrm{cm}^{-3}$, which
corresponds to $n_0=0.1 \mathrm{cm}^{-3}$.
Following Soker et~al. (\cite{soker}) we define the radius of a sphere which
contains an ISM mass equal to the shell mass
\begin{eqnarray}
\lefteqn{  R_0 \equiv \left( \frac {3\ M_s}{4 \pi \rho_0} \right)^{1/3} }
\nonumber\\
& & {}  = 3 \left( \frac {M_s}{0.2\ M_\odot} \right)^{1/3} 
  \left( \frac {n_0}{0.1\ \mathrm{cm}^{-3}} \right)^{-1/3} \mathrm{pc}.
  \label{eq:r0}
\end{eqnarray}
The stopping distance of the up-stream shell's segment is
given by (Eq.~5 of Soker et al. \cite{soker})
\begin{equation}
  r_{\rm stop} = R_0\ [2 \alpha\ (1+\alpha)]^{-1/3},
  \label{eq:dup}
\end{equation}
where $\alpha \equiv v_\ast/v_{\rm exp}$.
The time the up-stream segment reaches this maximum distance is
\begin{equation}
  t_{\rm stop} = \frac {R_0}{v_{\rm exp}} \alpha^{-2/3}.
  \label{eq:tup}
\end{equation}
As the down-stream segment expands undisturbed the diameter
of the shell along the stream at $t_{\rm stop}$ is
\begin{equation}
  d_{\rm stop} = R_0\ [(2 \alpha\ (1+\alpha))^{-1/3} + \alpha^{-2/3}]
  \label{eq:dstop}
\end{equation}
After reaching its maximum distance from the central star at
$r_{\rm stop}$, the up-stream segment of the shell is pushed by
the ISM toward the central star.

For an AGB star $v_{\rm exp} \simeq 10~\mathrm{km}~\mathrm{s}^{-1}$. 
The typical star-ISM
velocity at $\sim 150$~pc from the galactic plane
is $v_\ast > 10~\mathrm{km}~\mathrm{s}^{-1}$.
For $\alpha=1$ and assuming $M_s = 0.1~M_{\sun}$,
we get from Eqs. (\ref{eq:r0}), (\ref{eq:dup}) and (\ref{eq:dstop})
$r_{\rm stop}=1.6$~pc and $d_{\rm stop} = 4$~pc.
Even for an extreme case of $n_0=0.01~\mathrm{cm}^{-1}$,
$M_s = 1~M_{\sun}$ (which is a generous upper limit for
an AGB shell, especially if it were ejected in a thermal pulse from a low
mass star, as suggested in \cite{lers})
and $\alpha=0.25$ (say, expansion velocity of
$20~\mathrm{km}~\mathrm{s}^{-1}$
and $v_\ast = 5~\mathrm{km}~\mathrm{s}^{-1}$), we find $r_{\rm stop}=13.5$~pc
and $d_{\rm stop} = 43$~pc.
Given the observed diameter of the CO shell of $\sim$140~pc we can conclude
that an AGB shell would have been seriously disturbed by the ISM before reaching
these dimensions, its up-stream part would have to be significantly brighter
(because of significant accretion of the matter from the ISM)
than the opposite part and the central star would very likely be now observed
outside the up-stream rim. All this is not observed.

The above estimates are supported by observations of planetary nebulae.
From Eq. ({\ref{eq:r0}}) we see that when a typical planetary
nebula shell of $0.2\ M_{\sun}$ 
reaches a radius of $\sim 3$~pc,
it is expected to be highly distorted by the ISM.
Indeed, examining the list of planetary nebulae interacting with the ISM
compiled by Tweedy \& Kwitter (\cite{tweed}, their Table~3), we find that all
the planetary nebulae in this list have radii $< 5$~pc.
Most of the large planetary nebulae are highly distorted; not is only the central
star not at the center, but the shells are neither
circular nor elliptical.
Some planetary nebulae, like NGC\,6826
NGC\,2899 and A\,58 (surrounding the final helium shell flash star V605~Aql),
have very large, diameters $\sim 10-40~$pc, IRAS structures around them
(Weinberger \& Aryal \cite{wein}, Clayton \& De Marco \cite{clay}).
They are all severely distorted. Clayton \& De Marco (\cite{clay}) argue that
structures of this size are
swept-up ISM dust, rather than AGB mass-loss shells.

The above analysis rises a question: are the shells claimed
in \cite{lers} indeed real and related to V838~Mon?

It is difficult
to discuss the nature of the emission seen in the MSX image as it has been
recorded only in the A band (8.3~$\mu$m) image. No emission is seen in the
bands C (12.1~$\mu$m), D (14.6~$\mu$m) and E (21.4~$\mu$m). This is
perhaps due to the highest sensitivity of band A.

The elliptical structure around V838~Mon in the IRAS image shown in Fig.~1
of \cite{lers} at first sight looks convincing.
However, in the whole field of
this figure it is easy to fit several ellipses of similar sizes and similar
orientations as the one drawn by \cite{lers}. One possible explanation is that
the image pattern seen in Fig.~1 of \cite{lers} might be spurious, i.e. of
instrumental and/or image processing origin. Another likely interpretation is that
this is a general pattern of the interstellar diffuse emission in this region
and thus it has nothing to do with V838~Mon. The discussed region is a part of
an extended infrared emission related to several molecular clouds and HII regions
near the direction to V838~Mon (see Sect.~\ref{ism}).
Whether or not a part of this emission is
physically related to V838~Mon is an important question but it cannot be decided
just from the image.

Fig.~3 of \cite{lers} has been derived from a compilation of
surveys in the CO 1-0 line made by Dame et~al. (\cite{dame01}). However, if one
takes the composite map from Fig.~2 of Dame et~al. and expands near the
position of V838~Mon the resultant image is not the same as that in \cite{lers}.
In particular, there is no emission to the left and upper-left of the position
of V838~Mon [i.e. $l \ga l$(V838~Mon) and $b \ga b$(V838~Mon)], so no bubble-like
structure is seen. Apparently Dame et~al. (\cite{dame01}) considered this emission
as statistically insignificant. Indeed, the data for this region come from
a low resolution (0$\fdg5$) and low sensitivity survey of Dame et~al.
(\cite{dame87}). Thus the emissions seen in Fig.~3 of \cite{lers} and coming
from this last survey is uncertain and might be spurious.
This is supported by the fact that
two large, faint patches seen to the left and upper-left of the V838~Mon bubble
in Fig.~3 of \cite{lers} (not seen in Fig.~2 of Dame et~al. \cite{dame01})
cannot be identified with any known CO cloud
while two known CO regions, namely regions (9) and (13) in
Table~\ref{ism_tab} (see Sect.~\ref{ism}), are not seen in the image of \cite{lers}.

\subsection{Interstellar medium  \label{ism}}

The presence of the light echo proves that there is dusty matter around
V838~Mon. As argued in Tylenda (\cite{tyl}) and in Sect.~\ref{echo} of the
present paper this matter is likely to be of interstellar character. This notion
is supported by the conclusion of Sect.~\ref{phot} that V838~Mon is likely to be
a young binary system, as well as
by the fact that having the Galactic coordinates,
$l = 217\fdg80, b = +1\fdg05$, the object is located near the Galactic plane.
Therefore it is important to investigate observational data on the ISM in the
vicinity of V838~Mon. In this section we discuss the available data from the IRAS
and CO surveys.

\begin{table*}
\caption{Interstellar regions within $\sim 1\fdg5$ from the position of
         V838~Mon}
\label{ism_tab}
\begin{tabular}{r l l r l l l}
\hline
 No. & ~~~~$l$~~~~~$b$& Name          &  $T^{\ast}_A$ & $V_\mathrm{LSR}$
& Notes & References
\\
     & ~~~($\degr$)~~~($\degr$)&      &  (K)  & (km/s)    &       &           \\
\hline
  1 &  216.5\,+0.5  &                &       & 47.  & MC         & MAB \\
  2 &  216.5\,+1.2  &                &       & 48.  & MC         & MAB \\
  3 &  217.0\,+0.9  & FT\,84         &       & 51.  & MC         & Av, DHT  \\
  4 &  217.2\,+0.5  &                &  2.6  & 52.  & MC         & Av, WB, MAB \\
  5 &  217.3\,+0.0  &                &       & 27.  & MC         & MAB \\
  6 &  217.4\,$-$0.1& FT\,87, BFS\,57& 18.9  & 26.  & MC         & Av, WB \\
  7 &  217.4\,+0.3  & FT\,88, BFS\,58&  3.7  & 50.  & MC         & Av, WB, MAB \\
  8 &  217.6\,$-$0.2& FT\,89, BFS\,59& 10.4  & 26.  & MC         & Av, WB \\
  9 &  217.6\,+2.4  &                &  6.3  & 55.  & MC         & Av, WB \\
 10 &  217.9\,+0.9  &                &       &      & RN         & Ma \\
 11 &  218.0\,+0.2  &                &       & 23.  & MC         & MAB \\
 12 &  218.1\,$-$0.3& S\,287, FT\,91 & 18.2  & 26.--30.& HII, MC & Av, WB, MAB \\
 13 &  218.7\,+1.8  & IC\,466, S\,288&  6.3  & 57.  & HII, MC    & Av, WB \\
\hline
\end{tabular}
\\
\begin{flushleft}
Notes: HII -- HII region, MC -- molecular cloud, RN -- reflection nebula.\\
References: Av -- Avedisova (\cite{aved}), DHT -- $V_\mathrm{LSR}$ estimated from
original data of Dame et~al. (\cite{dame01}),
Ma - Magakian (\cite{mag}),
MAB -- May et~al. (\cite{may}), WB -- Wouterloot \& Brand (\cite{wout}).
\end{flushleft}
\end{table*}

\begin{figure}
\centering
  \resizebox{\hsize}{!}{\includegraphics{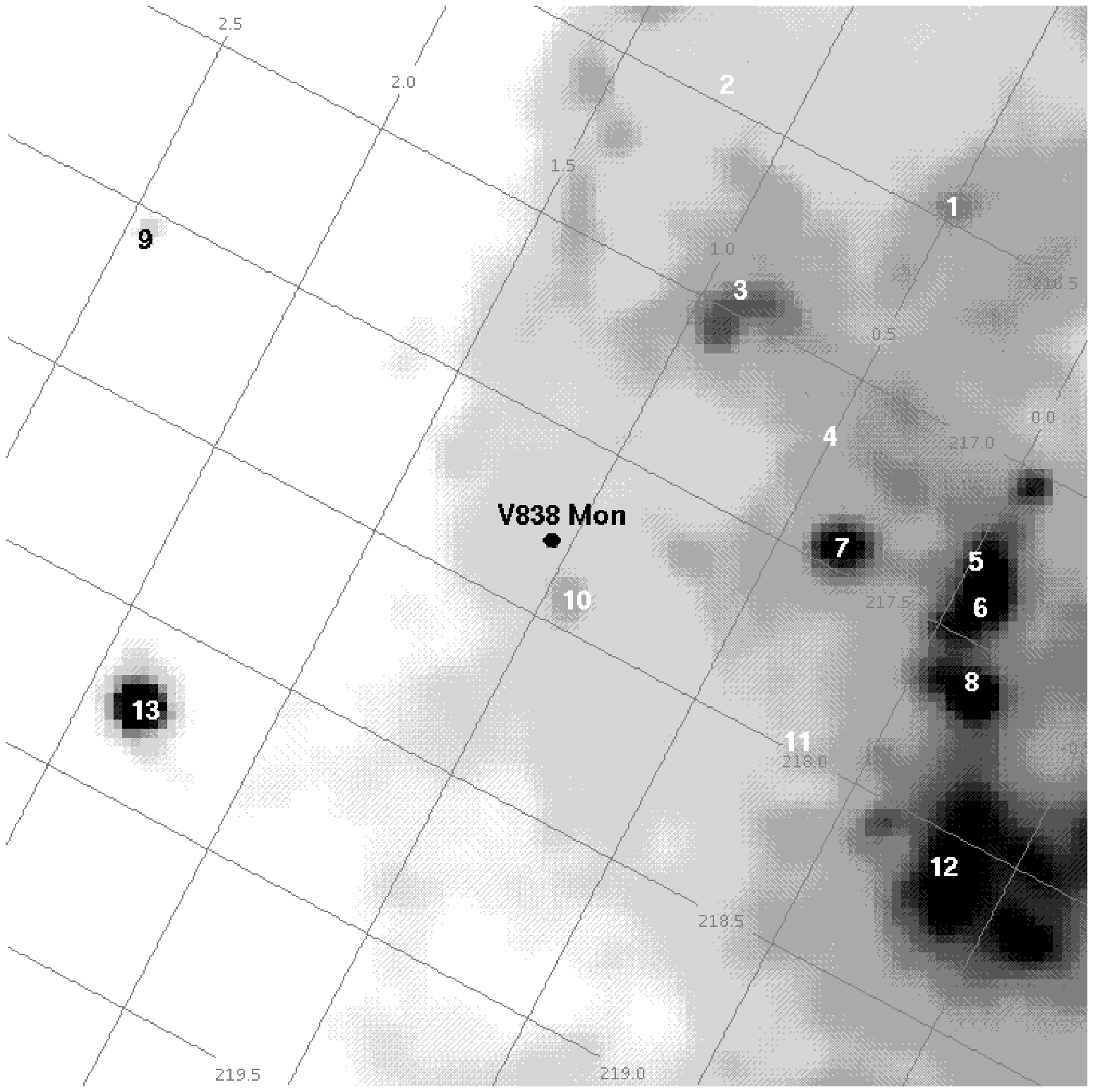}}
  \caption{The 100~$\mu$m IRAS image centered on the position of V838~Mon
(north is up, east to the left). The grid shows the galactic ($l,b$) coordinates.
Numbers show the positions of the ISM regions from Table~\ref{ism_tab}.
}
  \label{iras_f}
\end{figure}

Figure \ref{iras_f} shows the IRAS image at 100~$\mu$m centered at the position
of V838~Mon. At this wavelength practically all emission seen comes from
dust (interstellar or circumstellar). As can be seen from the figure
V838~Mon is located in a faint diffuse emission probably related to bright
regions near the Galactic plane.
The numbers in the figure show the positions
of molecular and HII regions listed
in Table~\ref{ism_tab}, which are within $\sim 1\fdg5$ of the position
of V838~Mon. The table
gives the galactic coordinates of the regions, their usual names, values of
$T^*_A$ taken from Wouterloot \& Brand (\cite{wout}), $V_\mathrm{LSR}$ resulting from
CO line observations, and types of the regions.
The references to the data are given in the last column of the table.

As can be seen from Table~\ref{ism_tab} the regions marked in Fig.~\ref{iras_f}
can be divided into two groups from the point of view of their positions, brightness
and $V_\mathrm{LSR}$. The brightest regions near and slightly
below the Galactic plane (regions 5, 6, 8, 12) have $V_\mathrm{LSR}$ in the range
of 20--30~km~s$^{-1}$.
The heliocentric radial velocity of V838~Mon is not well known as it
has been estimated from
outburst spectra. The results are within 55--65~km~s$^{-1}$
(Kolev et~al. \cite{kolev}, Kipper at~al.
\cite{kipper}, M.~Miko\l{}ajewski -- private communication).
That of the B-type companion is $\sim 64$~km~s$^{-1}$ (T.~Tomov -- private
comminication). Using the
results of Dehnen \& Binney (\cite{debin}) this can be transformed to
$V_\mathrm{LSR}$ = 44--54~km~s$^{-1}$. Thus the above regions
are most probably at much smaller distances (2--3~kpc if interpreted with
the Galactic rotation curve of Brand \& Blitz \cite{bb}) than V838~Mon and they
are simply seen in front of the object.

However the regions lying above the Galactic plane (regions 1, 2, 3, 4, 7, 9, 13)
have $V_\mathrm{LSR}$ between 47--57~km~s$^{-1}$. When interpreted with the rotation curve
of Brand \& Blitz (\cite{bb}) their distances are in the range of 6--8~kpc. Thus
these regions are located much closer to V838~Mon and a physical relation between
one of them and the matter seen in the light echo is quite possible.

Region (10), whose apparent position is closest to that of V838~Mon,
has been listed in Magakian (\cite{mag}) as a reflection nebula related
to a 9~magnitude B9 star HD~53135 (LS~$-$03~15) estimated to be at
a distance of $\sim 2$~kpc (Vogt \cite{vogt}, Kaltcheva \& Hilditch
\cite{kaltch}). Thus this region is probably located well in front of V838~Mon.

Interstellar Na~I lines in the spectrum of V838~Mon during eruption showed
two components at heliocentric velocities of $\sim 37$ and $\sim 64$~km~s$^{-1}$
(Zwitter \& Munari \cite{zwimun}, Kolev et~al. \cite{kolev}, Kipper et~al.
\cite{kipper}). When transformed to $V_\mathrm{LSR}$ the figures become $\sim 26$
and $\sim 53$~km~s$^{-1}$. Thus both line components can be interpreted as due to ISM
related to the above two groups of the ISM regions.
Detailed observations of the vicinity of V838~Mon in molecular lines might be
important for discussing the nature of V838~Mon.

\section{Discussion and summary  \label{discuss}}

The goal of this paper is to use available
observational data on the progenitor and enviroment of V838 Mon
to better constraint the nature of its
eruption. 
Below we summarize and discuss our main findings and conclusions.

(1) {\it The nature of the progenitor.}
In Sec.~\ref{phot} we have analyzed the photometric data available
for the progenitor. Most likely the progenitor was
a young binary system consisting of two
intermediate mass (5--10~$M_{\sun}$) stars. V838~Mon itself was either
a main sequence star of similar mass as its B-type companion or a slightly
less massive pre-main-sequence star. The system is very wide as the B-type
companion observed today does not seem to be affected by the eruption.
From the maximum photospheric radius of V838~Mon during eruption
(Tylenda \cite{tyl05}) we can estimate that the separation of the components
is $> 3 \times 10^3\ R_{\sun}$ so the orbital period is $> 12$~years.
The B-type companion was probably not involved in the eruption of
V838~Mon, at least directly. The hypothesis of a young binary system
is also supported by the position of the object near the Galactic plane
and the conclusion of Sect.~\ref{enviroment} that V838~Mon is probably embedded in the ISM.

A less likely hypothesis is that the presently observed B-type companion
does not form a binary system with V838~Mon. In this case, other than
a B-type main sequence star, the progenitor could have been an A--B spectral type giant
evolving from the main sequence or a B-type post-AGB star. These two
possibilities however involve a short (giant in the Hertzsprung gap)
or very short (post-AGB) evolutionary phase. As it is an old object in
these cases it would not be expected to reside inside or close to
dense ISM regions.

We can safely excluded the possibility that before eruption
V838~Mon was of spectral type K--M, so it could not have been 
a typical RGB or AGB star.

(2) {\it The light echo and the Galactic enviroment of V838~Mon.}
In several studies the light echo was used to argue that
the light-reflecting dust was
expelled by V838 Mon in previous eruptions (e.g Bond et al. \cite{bond},
\cite{lers}).
As argued by Tylenda (\cite{tyl}), and discussed here in Sect.~\ref{echo},
the data strongly suggests that dust is of ISM origin.
The dust structure derived from the echo analysis does not show any
hint of spherical symmetry. On the contrary,
the outer boundary of the echoing dust in front of the object can be
approximated by a plane at a distance
of $\sim 3.5$~pc from V838 Mon and inclined at
an angle of $\sim 26\degr$ to the line of sight. The strongly asymmetric
dust-free region in the near vicinity of V838~Mon, inferred from
the central hole in the echo, is interpreted as produced by the V838~Mon
progenitor (and possibly its B-type companion) moving relative to
the local ISM and sweeping out the medium by its fast wind.
As discussed in Sect.~\ref{ism}, there are several interstellar
molecular regions seen in the IRAS image and CO surveys probably located
near V838~Mon. The local
ISM seen in the light echo is therefore likely to be related to one or
some of them.

\acknowledgement{This has partly been supported
from a grant no. 2~P03D~002~25 financed by the Polish State Committee for
Scientific Research, as well as by the Israel Science Foundation.
We thank T. M. Dame for providing his CO data for a field centered on V838~Mon.}

\end{document}